\documentclass[12pt]{iopart}

\usepackage{graphicx}

\begin{document}

\graphicspath{{figures/}}

\article[]{}{J/$\psi$ production at forward rapidity in Pb-Pb collisions at ${\bf \sqrt{\mbox{\boldmath $s$}_{NN}}=2.76}$~TeV, measured with the ALICE detector}

\author{P. Pillot, for the ALICE Collaboration}
\address{Subatech (Universit\'e de Nantes, Ecole des Mines and CNRS/IN2P3), Nantes, France}
\ead{pillot@subatech.in2p3.fr}

\begin{abstract}
In the ALICE experiment, at forward rapidity ($2.5 < y < 4$), the production of heavy quarkonium states is measured via their $\mu^+\mu^-$ decay channels. We present the first measurement of inclusive J/$\psi$ production, down to $p_{\rm T} = 0$, from Pb-Pb data collected at the LHC at $\sqrt{s_{\rm NN}}=2.76$~TeV. Preliminary results on the nuclear modification factor ($R_{\rm AA}$) and the central to peripheral nuclear modification factor ($R_{\rm CP}$) show J/$\psi$ suppression with no significant centrality dependence and an integrated $R_{\rm AA}^{0-80\%}=0.49\pm0.03({\rm stat.})\pm0.11(\rm syst.)$.
\end{abstract}

Heavy quarkonium states have long been proposed as a sensitive probe of the strongly-interacting deconfined medium expected to be formed in the early stages of high-energy heavy-ion collisions~\cite{Satz86}. In particular, a measurement of the \mbox{in-medium} dissociation probability of different quarkonium states is expected to provide an estimate of the initial temperature of the medium. Suppression of J/$\psi$ production beyond that expected from cold nuclear matter (CNM) effects (nuclear absorption, \mbox{(anti-)shadowing, ...)} has indeed been observed at SPS and RHIC energies~\cite{NA60, PHENIX07, PHENIX11}, but several questions are left open. At the LHC, larger suppression might be expected, due to the higher initial temperature. However, according to regeneration scenarios~\cite{PBM00}, with $\sim$~10 times more $c\overline c$ pairs produced in central Pb-Pb collisions compared to RHIC, an enhancement of the J/$\psi$ production could also be observed. In the ALICE experiment, at forward rapidity ($2.5 < y < 4$), the inclusive production of heavy quarkonium states is measured down to $p_{\rm T} = 0$ via their $\mu^+\mu^-$ decay channels in the muon spectrometer, as describe in~\cite{ALICEpp}.

In fall 2010, the LHC delivered the first Pb-Pb collisions at a center of mass energy $\sqrt{s_{\rm NN}}=2.76$~TeV. ALICE collected data with a minimum bias (MB) trigger, defined as the logical AND between signals from the pixel detector (SPD) covering the range $|\eta|<2$ and the two scintillator arrays of the VZERO detector covering the ranges $2.8<\eta<5.1$ and $-3.7<\eta<-1.7$, in coincidence with two beam pick-up counters, one on each side of the interaction region. The centrality of the collision has been determined from the amplitude of the VZERO signal~\cite{Alberica}. The total data sample available for physics analysis amounts to $17\cdot10^6$ MB events. Additional cuts have then been applied to improve the purity of the muon sample, in particular by requiring both particles in the pair to be detected in the muon trigger stations. Furthermore, we perform the cuts $-4 < \eta < -2.5$ and $17.6 < R_{\rm abs} < 89$~cm, where $R_{\rm abs}$ is the radial coordinate of the track at the end of the hadronic absorber located in front of the spectrometer, to select muons in the geometrical acceptance of the detector.

\begin{figure}[htbp]
\includegraphics[width=7.8cm]{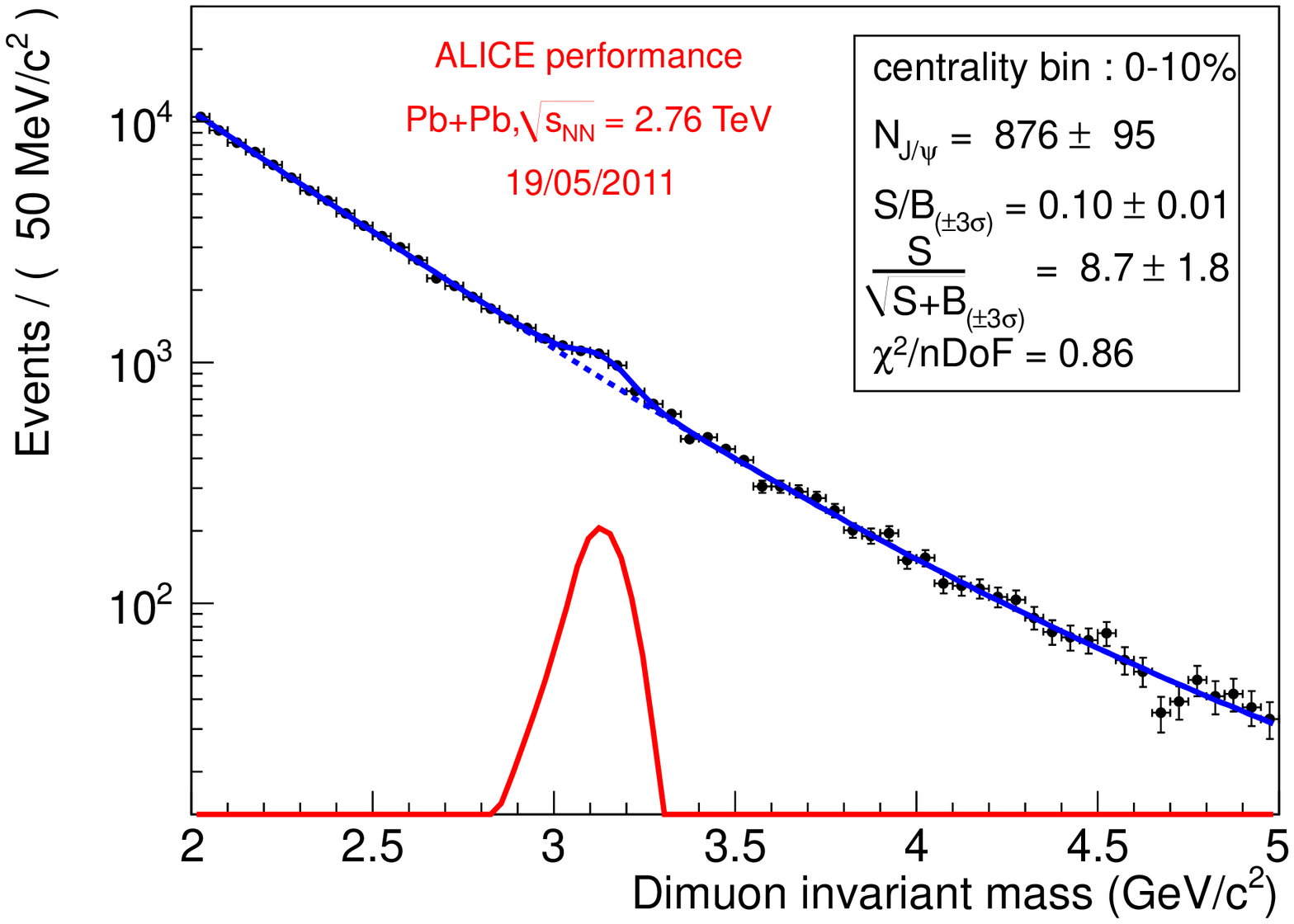}
\includegraphics[width=7.8cm]{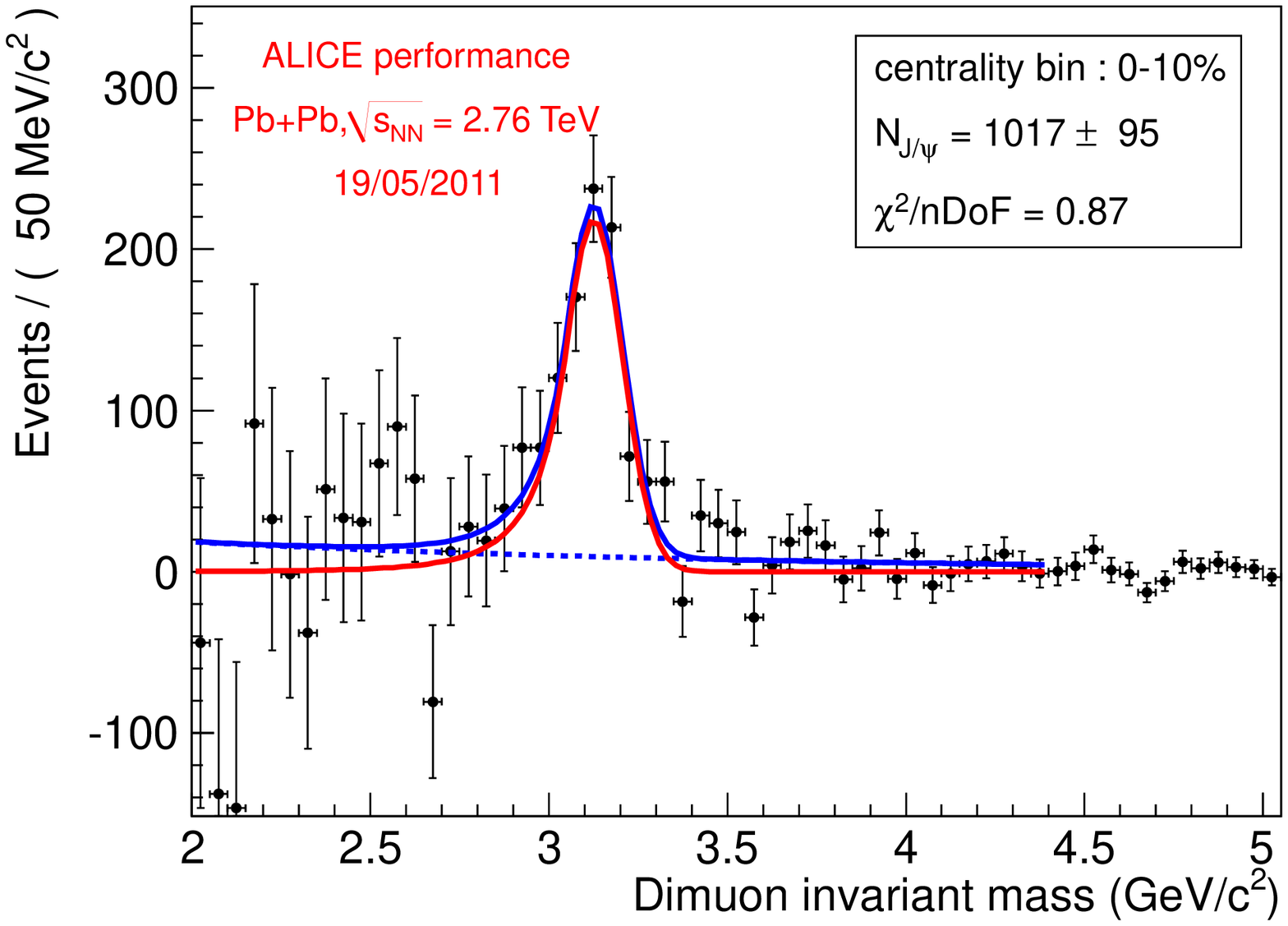}
\caption{\label{fig:mass}Invariant mass distribution for opposite-sign muon pairs in the centrality class 0-10\% before (left) and after (right) mixed-event combinatorial-background subtraction, with the result of the corresponding fits.}
\end{figure}

After these selections, the data sample has been divided into four centrality classes: \mbox{0--10\%}, 10--20\%, 20--40\% and 40--80\% of the inelastic Pb-Pb cross section. In each of these sub-samples, the number of J/$\psi$ particles has been extracted by using two different methods. We have first fitted the opposite-sign dimuon invariant mass distribution with a Crystal Ball (CB) function (a gaussian with polynomial tails) to reproduce the J/$\psi$ shape and a sum of two exponentials to describe the underlying continuum (Fig.~\mbox{\ref{fig:mass}-left}). Alternatively, we have subtracted the combinatorial background using the event-mixing technique and fitted the resulting mass distribution with a CB function and an exponential or a straight line to describe the remaining background (Fig.~\ref{fig:mass}-right). The width of the J/$\psi$ mass peak depends on the resolution of the spectrometer which could, $a~priori$, be affected by the detector occupancy which increases with centrality. This effect has been evaluated by embedding simulated J/$\psi$ decays to muons into real events and no modification has been observed. Such a conclusion has been confirmed by a direct measurement of the tracking chamber resolution versus centrality using reconstructed tracks. Therefore, the same CB line shape can be used for all centrality classes. Several parameters have been tested for the CB tails, either fixing them to the values obtained in p-p collisions (where the signal over background ratio is more favorable) or in simulations. For each of these choices, the mean and width of the gaussian part have been obtained by fitting the integrated mass distribution in the centrality range \mbox{0--80\%} and then varying the width by $\pm$ 1 standard deviation to account for uncertainties (varying the mean has turned out to have negligible effect in comparison). The number of J/$\psi$ particles in each centrality class $i$ ($N_{\rm J/\psi}^i$), as well as the ratios $N_{\rm J/\psi}^i / N_{\rm J/\psi}^{40-80\%}$, have been determined as the average of the results obtained with the two fitting approaches and the various CB parameterizations, while the corresponding systematic uncertainties have been defined as two times the RMS of these results, which also approximately corresponds to the maximum difference with respect to the mean value. The largest uncertainties have been obtained for the most central class and amount to 19\% and 12\% for $N_{\rm J/\psi}^{0-10\%}$ and $N_{\rm J/\psi}^{0-10\%} / N_{\rm J/\psi}^{40-80\%}$ respectively.

In order to extract the inclusive J/$\psi$ yield, $N_{\rm J/\psi}^i$ has been normalized to the number of MB events in the corresponding centrality class and further corrected for the branching ratio of the dimuon decay channel and the acceptance times efficiency ($A\times\epsilon$) of the detector. This latter quantity has been determined from  MC simulations. The generated J/$\psi$ $p_{\rm T}$ and $y$ distributions have been interpolated from existing measurements~\cite{Bossu}, including shadowing effects from EKS98 calculations~\cite{EKS}. The efficiencies of the muon trigger chambers have been measured directly from data then applied to the simulations. For the tracking apparatus, the time-dependent status of each electronic channel and their run-by-run evolution have been taken into account as well as the residual misalignment of the detection elements. We thus obtained a run-averaged value of $A\times\epsilon=19.4\%$, with a 7\% relative systematic uncertainty. The dependence of the tracking efficiency with the detector occupancy has also been evaluated using the embedding technique. A small decrease of this efficiency when increasing the centrality ($-2\%$ in the most central class) has been observed. This variation has been confirmed by a measurement of the tracking efficiency performed directly from data, and is included in the systematic uncertainties.

To measure the nuclear modification factors ($R_{\rm AA}^i$), the J/$\psi$ yield in the centrality class $i$ has been normalized to the inclusive J/$\psi$ cross-section measured in p-p collisions in the same rapidity domain at the same energy ($\sigma_{\rm J/\psi}^{\rm inclusive} = 3.46\pm0.13({\rm stat.})\pm0.32({\rm syst.})\pm0.28({\rm syst.lumi.})\mu {\rm b}$)~\cite{Roberta} and scaled by the corresponding nuclear overlap function ($T_{\rm AA}^i$) calculated using the Glauber model, while the ratios $N_{\rm J/\psi}^i / N_{\rm J/\psi}^{40-80\%}$ have been normalized to the ratios $T_{\rm AA}^i / T_{\rm AA}^{40-80\%}$ to extract the central to peripheral nuclear modification factors ($R_{\rm CP}^i$). The systematic uncertainties on the Glauber model calculations are 4\% and 6\% for $T_{\rm AA}^{0-10\%}$ and $T_{\rm AA}^{0-10\%} / T_{\rm AA}^{40-80\%}$ respectively.

\begin{figure}[htbp]
\includegraphics[width=7.9cm]{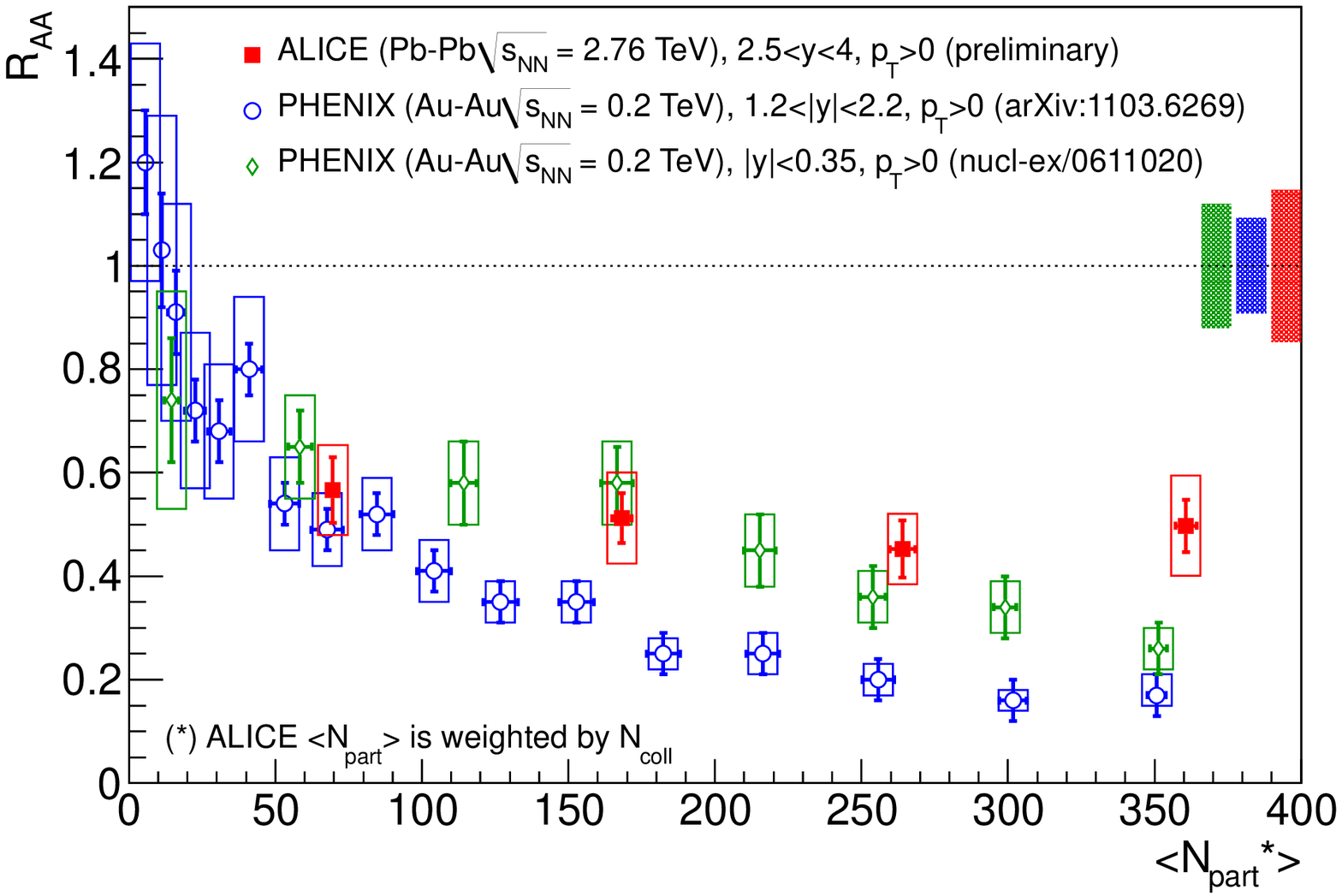}
\includegraphics[width=7.9cm]{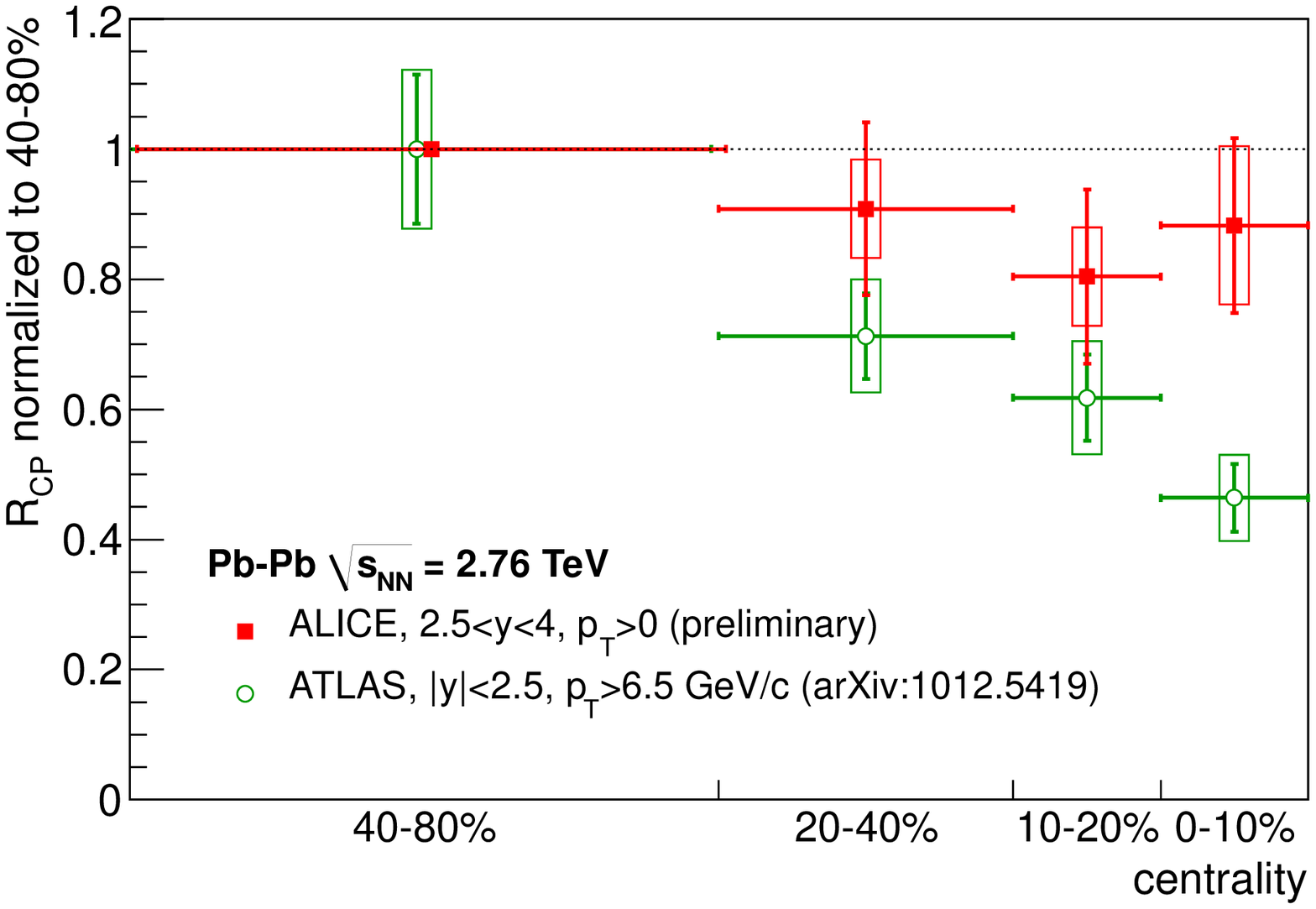}
\caption{\label{fig:RAARCP}Left: J/$\psi$ $R_{\rm AA}$ as a function of $\langle N_{\rm part} \rangle$ compared with PHENIX results in Au-Au collisions at $\sqrt{s_{\rm NN}}=200$~GeV. Right: J/$\psi$ $R_{\rm CP}$ as a function of centrality compared with ATLAS results. Error bars represent the statistical uncertainties, open boxes represent the centrality-dependent systematic uncertainties while the centrality independent uncertainties are shown by filled boxes.}
\end{figure}

The inclusive J/$\psi$ $R_{\rm AA}$ are shown in Fig.~\ref{fig:RAARCP}-left as a function of  the average number of nucleons participating to the collision ($\langle N_{\rm part} \rangle$) calculated using the Glauber model. To account for the bias due to our large centrality bins, $\langle N_{\rm part} \rangle$ has been weighted by the number of binary nucleon-nucleon collisions ($N_{\rm coll}$), which is expected to be the scaling variable of the J/$\psi$ production cross-section in A-A, in absence of nuclear matter effects. This correction is small, except for the most peripheral bin where $\langle N_{\rm part} \rangle$~=~46 while the weighted value is 70. These results show no significant dependence on centrality, and the integrated $R_{\rm AA}^{0-80\%}=0.49\pm0.03({\rm stat.})\pm0.11({\rm syst.})$. The comparison with the PHENIX measurements at $\sqrt{s_{\rm NN}}=200$~GeV~\cite{PHENIX07, PHENIX11} shows that the inclusive J/$\psi$ $R_{\rm AA}$ at 2.76~TeV in the rapidity domain $2.5 < y < 4$ are clearly above those measured at 200~GeV in $1.2 < |y| < 2.2$, while they are closer to the midrapidity values at 200~GeV (except in the most central collisions). The contribution from the B feed down to the J/$\psi$ production in our rapidity and $p_{\rm T}$ domain has been measured to be $\approx10\%$ in p-p collisions at $\sqrt{s_{\rm NN}}=7$~TeV~\cite{LHCb}. Therefore, the difference between the prompt J/$\psi$ $R_{\rm AA}$ and our inclusive measurement is expected to be $\approx11\%$ if the $b$ production scales with $N_{\rm coll}$ or smaller if it is suppressed by nuclear effects (shadowing, ...). Finally, the comparison of our J/$\psi$ $R_{\rm CP}$ results to the ATLAS measurements in the same centrality classes~\cite{ATLAS} \mbox{(Fig.~\ref{fig:RAARCP}-right)} indicates that the J/$\psi$ mesons measured at forward rapidity down to $p_{\rm T}=0$ are less suppressed than the high-$p_{\rm T}$ J/$\psi$ mesons at midrapidity (80\% of the J/$\psi$ particles measured by ATLAS have a $p_{\rm T}$ larger than 6.5 GeV/$c$).

In summary, these results show a significant suppression of the inclusive J/$\psi$ production in Pb-Pb collisions at $\sqrt{s_{\rm NN}}=2.76$~TeV. The comparison with PHENIX results suggests that re-generation mechanisms could play a role. In order to provide tight constraints to suppression/regeneration models, a better knowledge of CNM effects is required, which can be obtained with a measurement of J/$\psi$ production in p-A collisions at the LHC.

\end{document}